\def\rmin{r_{\rm min}}
\def\rmax{r_{\rm max}}

\def\pc{\;{\rm pc} }

\def\pc{\;{\rm pc} }

\def\Msun{\,{\rm M}_\odot}
\def\msun{\Msun}

\def\kms{\;{\rm km}~{\rm s}^{-1}  }

\def\Ds{{D_{\rm s}}}
\def\spose#1{\hbox to 0pt{#1\hss}}
\def\lta{\mathrel{\spose{\lower 3pt\hbox{$\sim$}}
    \raise 2.0pt\hbox{$<$}}}
\def\gta{\mathrel{\spose{\lower 3pt\hbox{$\sim$}}
    \raise 2.0pt\hbox{$>$}}}

\documentclass[12pt,preprint]{aastex}
\begin{document}

\title{Microlensing Maps for the Milky Way Galaxy}
\author{N. W. Evans and V. Belokurov}
\affil{Theoretical Physics, Department of Physics, 1 Keble Road,
Oxford, OX1 3NP, UK}
\begin{abstract} 
At any instant, there are $\sim 1000$ microlensing events to sources
brighter than 20th magnitude in the Milky Way Galaxy.  Large-scale
maps of the microlensing optical depth and the mean timescale are
constructed for a number of models of the Galactic bar and disk,
incorporating the effects of streaming and spiral
structure. Freudenreich's model can reproduce the high optical depths
towards the Bulge. It is also in good agreement with the data towards
the spiral arms (except for the field $\gamma$ Norma).  Spiral
structure tends to increase the optical depth by $\lta 20 \%$ and the
mean timescale by $\lta 100 \%$.  Different bar morphologies give
characteristically different shaped contours, especially at low
Galactic latitudes ($|b| < 2^\circ$). These could be traced out with a
K band microlensing survey, consuming $\sim 100$ minutes per night on
a telescope like VISTA.
\end{abstract}

\section{INTRODUCTION}
Microlensing surveys of the Galaxy are important because they
delineate the mass distribution directly.  The total number of events
identified by the OGLE and MACHO collaborations towards the Galactic
Bulge now exceeds a thousand (e.g., Wozniak et al. 2001; Alcock et
al. 2000). The identification of events towards spiral arms has been
reported by the EROS and OGLE collaborations (Mao 1999; Derue et
al. 2001).  Maps of optical depth as a function of Galactic latitude
and longitude were first drawn by Evans (1994).  Subsequently, a
number of authors emphasised the importance of exploiting information
on the spatial variation of the optical depth (e.g., Han \& Gould
1995; Zhao, Spergel \& Rich 1996; Gyuk 1999).

Microlensing searches are being done, or will be done in the near
future, in almost any longitude direction. The first of the next
generation experiments is the OGLE III venture (see
``http://sirius.astrouw.edu.pl/~ogle/'') which uses an 8k MOSAIC
camera with a field of view equal to $35' \times 35'$. The future
holds still greater promise with the advent of the new class of survey
telescopes like the VST (``http://www.eso.org/projects/vst/'') and
VISTA (``http://www.vista.ac.uk'').  VISTA has a field of view of 2.25
square degrees in the optical, while the VST has a smaller, but still
substantial, field of view of 1 square degree. So, there is a need for
large-scale maps of both the microlensing optical depth and the mean
timescale for the Galaxy.  These maps can be used to pick target
fields, to plan search methodologies for future experiments and to
assess what can be learnt about the structure of the Milky Way from
the distribution of microlensing events.

\section{MAPS OF THE GALAXY}

\begin{figure}
\begin{center} 
\plotone{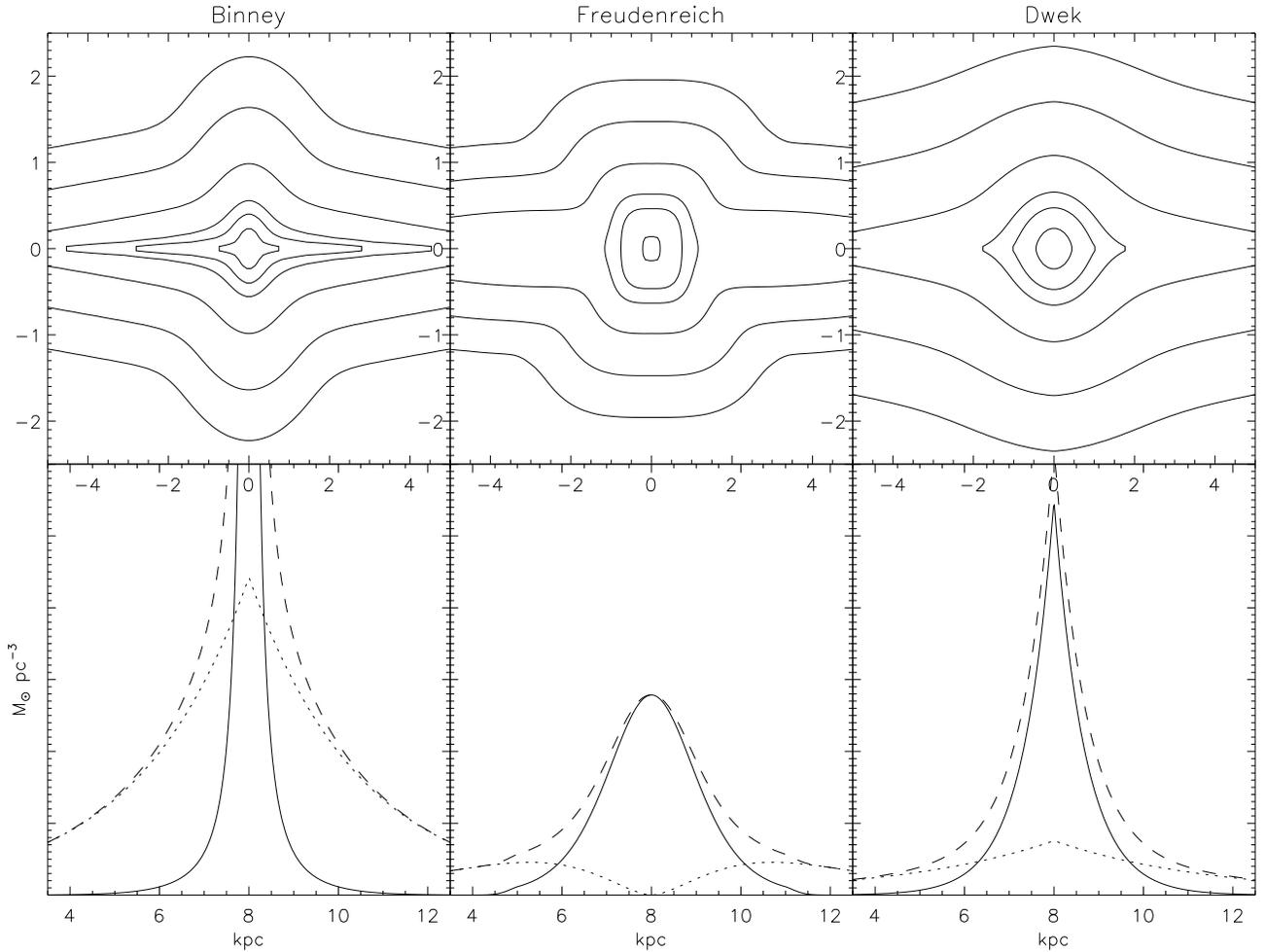}	
\end{center}
\caption{Upper panels show slices through the principal plane
of the bars of Binney et al., Freudenreich and Dwek et al. The
contour levels are ($10^{-3}, 10^{-2}, 10^{-1}, .5, 1, 2.5) \Msun
\pc^{-3}$. Lower panels show the density contributions of the bar 
(full line), disk (dotted line) and total (dashed line) along the line
of sight to the Galactic Center}
\label{fig:slice}
\end{figure}

We study three models of the inner Galaxy. The starting point of all
three models is the same, namely the infrared surface brightness maps
seen by the DIRBE instrument on the COBE satellite.  The first is the
model of Binney, Gerhard \& Spergel (1997), which is partially revised
in Bissantz et al. (1997). Here, the observed luminosity at $\sim 240$
microns, which is dominated by thermal emission from dust, is used to
deduce the three-dimensional spatial distribution of the dust. This
gives a short, flattened, cuspy bar with axis ratio $1:0.3:0.3$ and
with a viewing angle $\phi_0 = 20^\circ$. The second is the model of
Freudenreich (1998). Here, a mask of areas believed to be contaminated
with dust is constructed from maps of color variations.  The mask is
used to excise portions of the DIRBE data and the remainder of the
data is fitted. Note that the mask removes almost all of the data
within $|b| \lta 5^\circ$ for longitudes within $90^\circ$ of the
Galactic Center (see Figure 1 of Freudenreich's paper).  This gives an
extended, diffuse, swollen bar with axis ratio $1:0.37: 0.27$ and with
a viewing angle $\phi_0= 14^\circ$. The disk has a central hole with a
radius of $\sim 3$ kpc.  The third is the E2 model of Dwek et
al. (1995), as partially revised by Stanek et al. (1997).  The density
contours are stratified on concentric ellipses with axis ratio
$1:0.42:0.28$ with a viewing angle $\phi_0 =24^\circ$.  In correcting
for dust, Dwek et al. assumed a uniform foreground screen.  The bar is
less massive and less elongated than Freudenreich's and the disk does
not have a central hole.  To ensure a fair comparison, all three
models are normalized to have the same total mass of $1.5 \times
10^{10} \Msun$ within 2.5 kpc. Optical depths scale in rough
proportion with total mass, and so our results can be easily converted
to other preferred values. The three models are illustrated in
Figure~\ref{fig:slice}. Cuts through the principal plane of the bar
and along the line of sight to the Galactic Center are shown.
Table~\ref{table:wyn} shows the masses of the components in the three
models. Freudenreich's bar is the most massive, Binney et al.'s the
least massive.

EROS and OGLE are looking towards spiral arms. Accordingly, we
include the effects of spiral arms in our calculations. We assume that
the inner spiral pattern is two-armed and given by multiplying the
density by eqn (2) of Binney et al. (1997), namely
\begin{equation}
1 + \epsilon \cos^6 ( \phi - 0.95(r - \rmin) - \phi_0), \qquad
\epsilon = \tanh(r -\rmin).
\end{equation}
Here, $\rmin$ is proportional to the length of the bar and is 1.5 kpc
for Binney et al.'s bar, 2.25 kpc for Freudenreich's and 1.7 kpc for
Dwek et al.'s. This density factor is applied between $\rmin$ and
$\rmax = 3.5$ kpc.  The outer spiral pattern is four-armed and given
by $1 + 2\epsilon \cos^6 (2\phi - 1.1(r - \rmin) - 15^\circ)$ between
$\rmin =4.1$ kpc and $\rmax =8.5$ kpc.  The phase and the wavenumber
of the inner and outer spirals are chosen to match the longitudes of
the principal spiral arms given by Englmaier (2000). The arm/interarm
contrast is 2 for the inner and 3 for the outer spiral (cf. Rix \&
Rieke 1993). This is reasonable for a young population, but an
overestimate for old low mass stars that may make up the bulk of the
lensing population. Our results on the effects of spirality are upper
limits.

\begin{figure}
\begin{center} 
\plotone{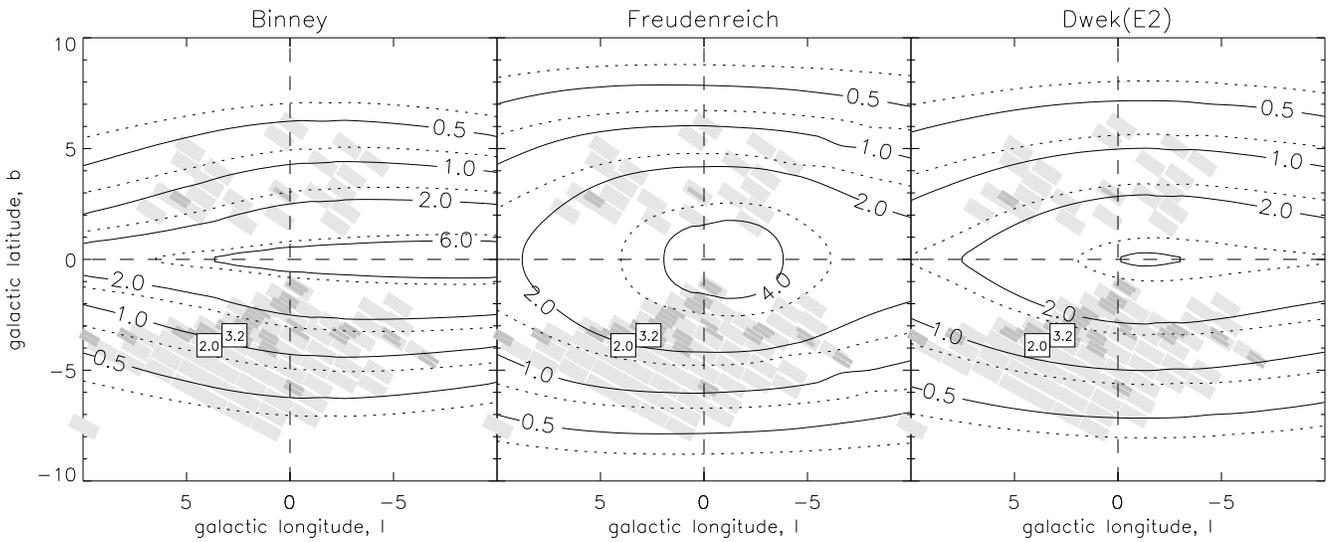}	
\end{center}
\caption{Contours of microlensing optical depth to the red clump 
giants (in units of $10^{-6}$) in the three bar models, excluding
(full lines) and including (dotted lines). The optical depths reported
by Alcock et al. (2000) and Popowski et al. (2000) are shown in
boxes. Light (dark) gray boxes correspond to EROS (OGLE II) fields.}
\label{fig:mmaps}
\end{figure}

Microlensing maps show contours of the source-averaged microlensing
optical depth $\langle \tau \rangle $ computed via (Kiraga \&
Paczy\'nski 1994)
\begin{equation}
\langle \tau \rangle = {\int_0^\infty \rho(\Ds) \tau(\Ds) 
                       \Ds^{2 + 2\beta} d\Ds \over
\int_0^\infty \rho(\Ds) \Ds^{2 + 2\beta} d\Ds},
\end{equation}
where $\Ds$ is the source distance and $\rho$ is the density of
deflectors.  For red clump stars, $\beta \approx 0$; for main sequence
stars, $\beta \approx -1$.  Red clumps stars are bright and
distinctive residents of the bar (e.g., Paczy\'nski \& Stanek 1998).
Figure~\ref{fig:mmaps} shows contours of optical depth to the red
clump in the three models. The dotted contours show the effect of
including the spiral structure. The amplification caused by spirality
varies according to the line of sight but it is typically $\sim 20\%$.
Of the three, Binney et al.'s gives the least symmetric microlensing
map, especially close to the Galactic plane where the gradients are
very steep.  Freudenreich's is the most symmetric, as it possesses the
smallest viewing angle.  The boxes mark the two locations where the
optical depth of bar stars has been measured. The first is $\ell =
3.9^\circ, b = -3.8^\circ$ where Popowski et al. (2000) report $2.0
\pm 0.4 \times 10^{-6}$ for the optical depth to the red clump
stars. The second is $\ell = 2.68^\circ, b = -3.35^\circ$ where Alcock
et al. (2000) report $3.2 \pm 0.5 \times 10^{-6}$ for the optical
depth to bar stars.  The value of the optical depth at $\ell =
3.9^\circ, b = -3.8^\circ$ contributed by each component in the three
models is recorded in Table~\ref{table:baade}.  Freudenreich's model
is in good agreement with the two measurements of optical depth to the
bar sources.

Even though all three models have the same total mass within 2.5 kpc,
Freudenreich's model has the higher optical depth because the bar is
more extended.  Notice that the shape of the contours in
Figure~\ref{fig:mmaps} becomes very similar for all three models once
$|b| > 4^\circ$. This makes it challenging to characterize the bar
morphology on the basis of data from the observed fields.  One qualm
is that Freudenreich excised most of the DIRBE data within $|b| \lta
5^\circ$ of the Galactic plane. In this region, his model is entirely
extrapolated from the light distribution at higher latitudes and so
may be unreliable. Binney et al.'s model reproduces the strong
concentration in the light near the Galactic plane. It lies within
$\approx 2 \sigma$ of Popowski et al.'s measurement when spiral
structure is included. Dwek et al.'s model with spiral structure is
just outside $1 \sigma$.  Can we modify Binney et al.'s model to
reproduce the microlensing optical depth data?  To get agreement with
the data, the bar needs to be made both longer and fatter.  For
example, changing the axis ratio to $1:0.6:0.4$ (as originally
envisaged in Binney et al. 1997) and increasing the total mass within
2.5 kpc by $50 \%$ gives a model in reasonably good agreement (namely
$\tau = 1.5 \times 10^{-6}$ at $\ell = 3.9^\circ, b = -3.8^\circ$
excluding spiral structure; this rises to $1.8 \times 10^{-6}$ when
spiral structure is included).

\begin{figure}
\begin{center} 
\plotone{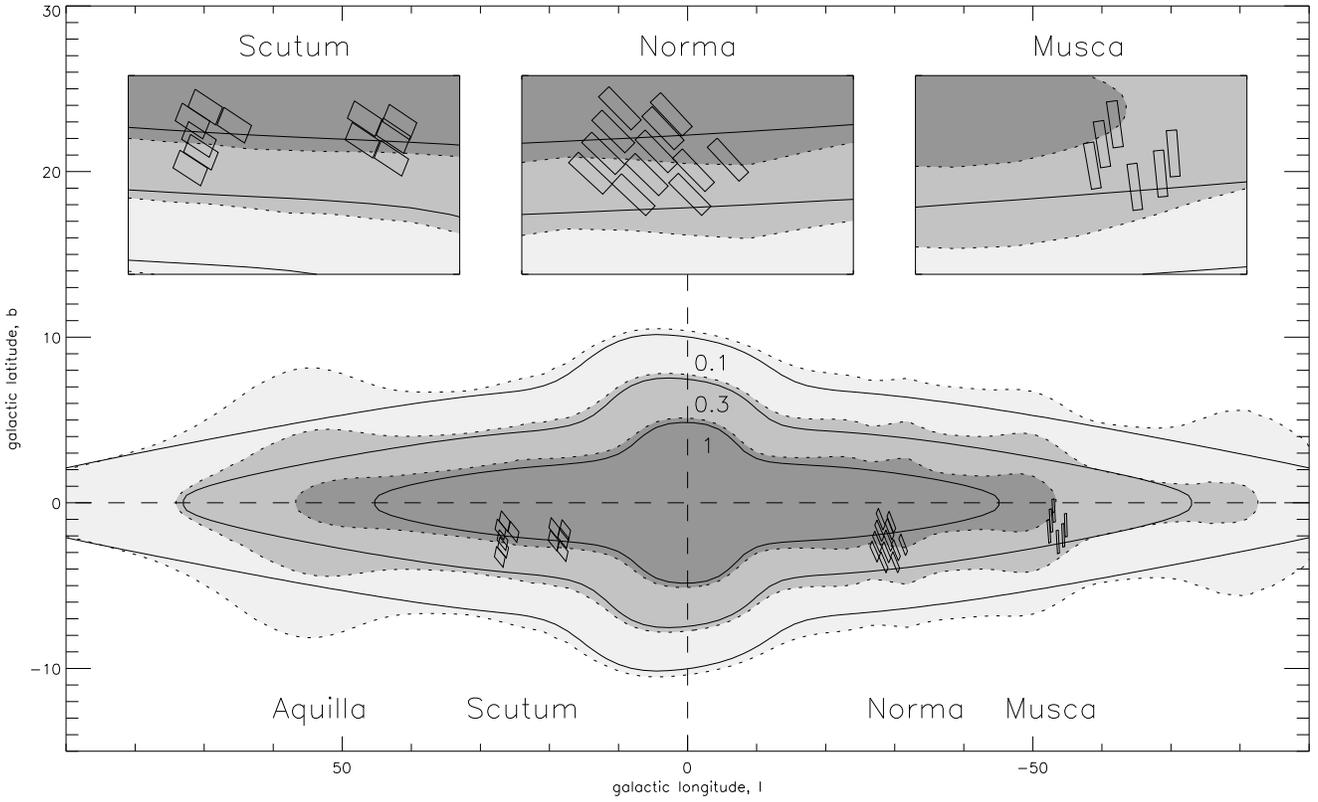}	
\end{center}
\caption{Contours of microlensing optical depth to all sources
(in units of $10^{-6}$) in Freudenreich's model, excluding (full
lines) and including (dotted lines) spiral structure. The insets show
the details of EROS fields towards the Scutum, Norma and Musca spiral
arms.}
\label{fig:bigmap}
\end{figure}

Figure~\ref{fig:bigmap} shows the contours of optical depth to all
sources in the inner Galaxy using Freudenreich's bar. The Galactic
disk has a sech-squared vertical profile with scaleheight 167 pc and
an exponential horizontal profile with scalelength 2.5 kpc.  The
insets show the details of three areas towards the Scutum, Norma and
Musca spiral arms that are being monitored by the EROS group. We see
that there is a factor of $\sim 6$ variation in the optical depth
across the EROS fields towards all three spiral arms. Freudenreich's
bar is so distended that it causes a thickening of the optical depth
contours even at longitudes well away from the center.  We can exploit
the maps to estimate the number of ongoing microlensing events of
stars brighter than 20th magnitude in the Milky Way.  At any instant,
there are $\sim 1000$ microlensing events, where we have allowed for
extinction in the same manner as Belokurov \& Evans (2002).
Table~\ref{table:spiral} compares the predictions of the three models
with the data provided by EROS. Notice that Freudenreich's model is in
excellent agreement with the data, except for the field $\gamma$
Norma, where the optical depth is about $2 \sigma$ away from the data.

\section{TIMESCALE MAPS}

It is also interesting to build maps of the mean Einstein crossing
timescale.  We assume that the mass function of the bar (or disk) is a
power-law between $0.01 \msun$ and $0.5 \msun$ with index $-1.33$ (or
$-0.54$) as suggested by Zoccali et al. (2001).  Let ($x,y,z$) define
coordinates along the major, intermediate and minor axis of the bar.
We calculate the average velocity dispersions required to reproduce
the shape of Freudenreich's bar guided by the tensor virial theorem
(e.g., Han \& Gould 1995, Blum 1996). We assume that stars on the
front side of the bar move along the major axis with a streaming
velocity of $\langle v_x \rangle = 50 \kms$, while stars on the back
side move with $\langle v_x \rangle = -50 \kms$ and we adjust the
dispersion $\sigma_x$ to preserve the total kinetic energy required by
the virial theorem. With these assumptions, the velocity distribution
is Gaussian about the streaming velocity with dispersions $\sigma_x =
100 \kms$, $\sigma_y = 80 \kms$, $\sigma_z = 68 \kms$.  For the disk
lenses, the random component has $\sigma_R = 34 \kms$, $\sigma_\phi =
21 \kms$, $\sigma_z = 18 \kms$ about a mean velocity $\langle v_\phi
\rangle $ of $214 \kms$ (see e.g., Edvardsson et al. 1993; Belokurov
\& Evans 2002). Although small velocity perturbations will be
associated with the spiral arms, these are neglected in our
calculations.

\begin{figure}
\begin{center} 
\plotone{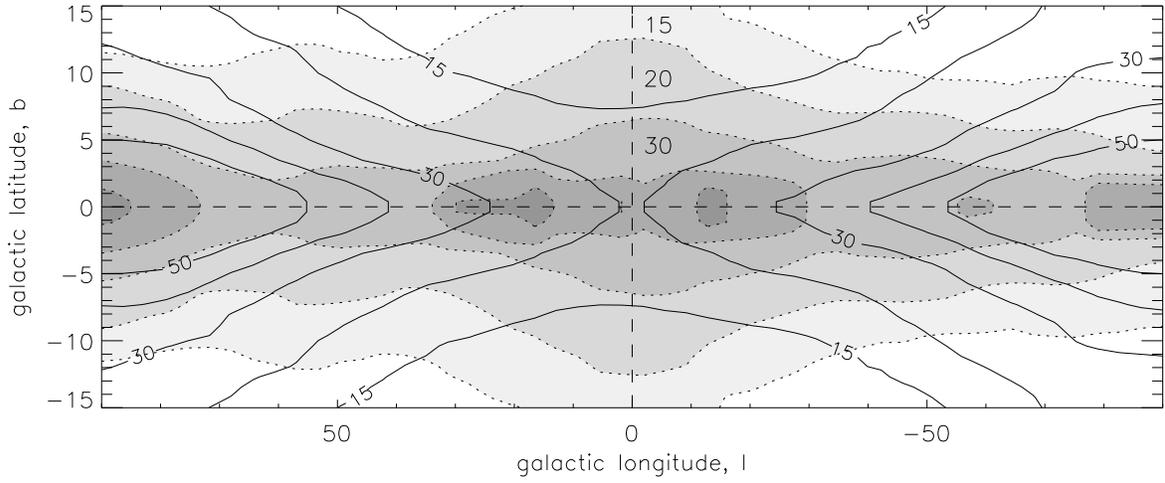}	
\end{center}
\caption{Contours of mean timescale (in days) for microlensing events
with sources and lenses lying in either the disk or Freudenreich's
bar. The timescale is the time taken to cross the Einstein radius.
The full lines exclude the contribution of spiral structure, the
dotted lines include the contribution. The contour levels are at
$15,20,30,40$ and 50 days.}
\label{fig:mapte_one}
\end{figure}

\begin{figure}
\begin{center} 
\plotone{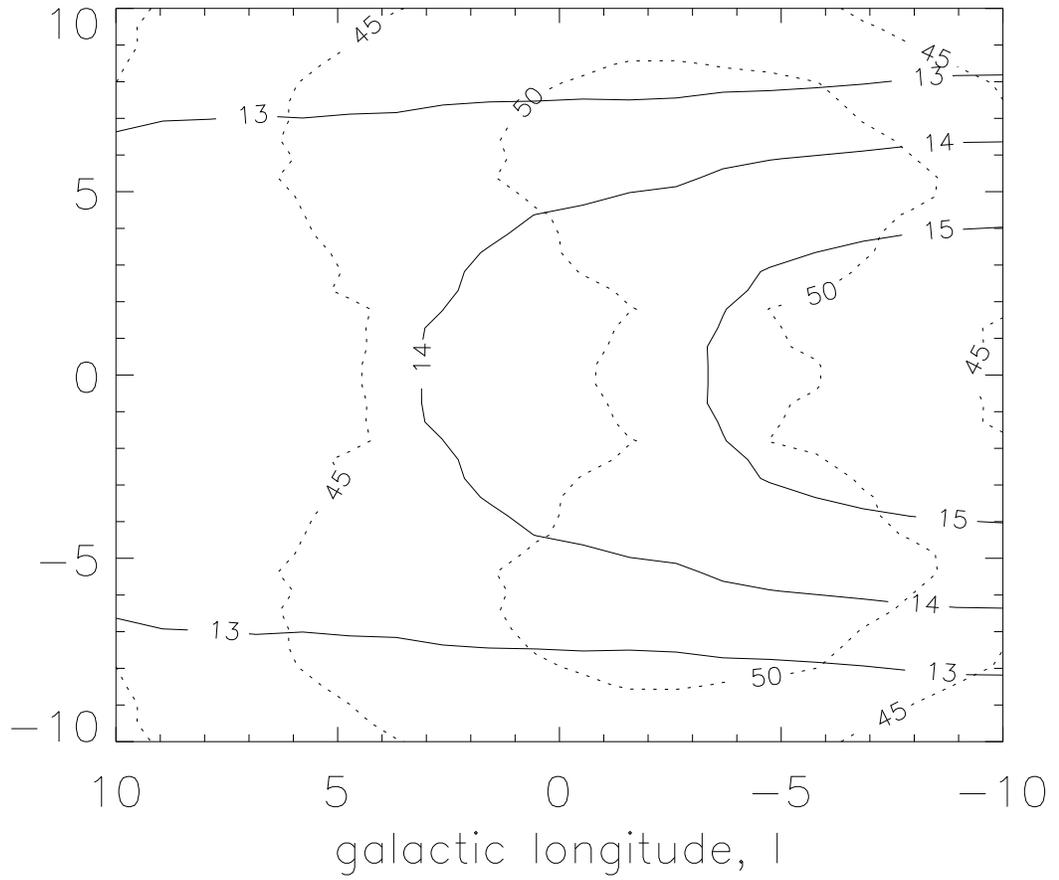}	
\end{center}
\caption{Contours of mean timescale (in days). The sources lie in the 
bar alone (Freudenreich's model), the lenses may lie in either the
disk or the bar. The full lines exclude the contribution of bar
streaming, the dotted lines include the contribution.}
\label{fig:mapte_two}
\end{figure}

Figure~\ref{fig:mapte_one} shows contours of the mean timescale for
events with sources and lenses in either the disk or Freudenreich's
bar, including (dotted lines) and excluding (full lines) the spiral
structure. The mean timescale is shortest towards the Galactic Center
and becomes longer at increasing Galactic longitudes. This is easy to
understand because the motion of the lens is directed more and more
along the line of sight at larger longitudes and so the transverse
velocity is typically smaller.  Notice that spiral structure has a
dramatic effect on the mean timescale.  For example, at Baade's
Window, the mean timescale is increased by a factor of $\sim 100 \%$
on incorporating the effects of spirality.  Figure~\ref{fig:mapte_two}
shows the inner $20^\circ \times 20^\circ$ in Freudenreich's model
including (dotted lines) and excluding (full lines) the effects of bar
streaming. This detail is drawn for sources in the bar only, and so
the asymmetry in the map is substantial.  Streaming is an important
effect because it removes kinetic energy from random motions and
places it in systematic motions that are directed almost along the
line of sight. There is a increasing gradient in the mean timescale
from the near-side to the far-side of the bar. This is a geometric
effect, in that lines of sight to the near-side are more nearly
perpendicular to the major axis than lines of sight to the
far-side. Popowski et al. (2000) report that about 40 \% of the
optical depth is in events with timescales $> 50$ days and that this
is at odds with standard Galactic models. By contrast, we find that
long timescale events are to be expected when streaming is taken into
account in the modelling.

\section{CONCLUSIONS}

We have drawn large scale maps of the optical depth and timescale
distribution for microlensing in the Galaxy.  Freudenreich's bar does
give a reasonable representation of the microlensing data to the
Bulge.  It recovers the optical depth towards the spiral arms (with
the exception of $\gamma$ Norma).  However, Freudenreich's model is
hollow. Stars otherwise expected to be in the central parts of the
disk can instead be used to augment the bar where they are efficient
at microlensing.

As pointed out by Gould (1995), microlensing surveys in the K band
would be very valuable to distinguish between models. This is all the
more true given the capabilities of the new generation of survey
telescopes.  VISTA has a field of view of 0.5 square degrees in the K
band.  Assuming that the seeing is $0.8''$ in Chile and scaling the
results of Gould (1995), then we estimate that VISTA will monitor
$\sim 1.5 \times 10^6$ stars in a single field of view for
crowding-limited K band images towards the Bulge. This means that we
are probing the luminosity function down to $K \sim 16$, assuming 3
magnitudes of extinction.  Photometry accurate to $3 \%$ for a $K \sim
16$ star will take $\sim 1$ minute on VISTA. Hence, a K band survey of
a $5^\circ \times 5^\circ$ field close to the Galactic Center will
take $\sim 100$ minutes every night, allowing 50 minutes for readout,
slew and guide star acquisition time. So, a K band microlensing survey
of the inner Galaxy is an attractive and feasible proposition with
VISTA.

\acknowledgments
NWE is supported by the Royal Society and VB acknowledges financial
help from the Dulverton Fund. We thank Bohdan Paczy\'nski and Shude
Mao for emphasizing to us the importance of large-scale maps, as well
as James Binney for discussions on the structure of the Galactic bar.
The anonymous referee is thanked for a helpful report.

\clearpage
\clearpage
\begin{table}
\begin{tabular}{l|cccccc}\hline
Model & bar & disk & bar+disk & disk & inner spiral & outer
 spiral\\ & $<2.5$ kpc & $<2.5$ kpc & $<2.5$ kpc & $<8.5$ kpc 
& 3.5 kpc & 8.5 kpc\\
\hline
Binney et al& 0.5 & 1.0 & 1.5 & 3.2 & 0.3 & 0.9 \\
Freudenreich & 1.1 & 0.4 & 1.5 & 3.2 & 0.2 & 1.2 \\
Dwek et al& 0.9 & 0.6 & 1.5 & 2.6 & 0.2 & 0.9 \\
\hline
\end{tabular}
\caption{This table shows the mass (in units of $10^{10} \Msun$) in
different components of the Galaxy in the three models. The fifth
column is the additional mass within 2.5 kpc caused by the inner
spiral alone.  The sixth column is the additional mass within 8.5 kpc
caused by the outer spiral alone.}
\label{table:wyn}
\end{table}
\clearpage
\begin{table}
\begin{center}
\begin{tabular}{l|cccccc}\hline
source & bar & bar & bar & bar+disk & bar+disk & disk \\
lens & bar & disk & bar+disk & bar+disk & disk & disk\\
\hline
Binney et al& 0.3 & 0.6 (0.9) & 0.9 (1.2) & 0.7 (0.8) & 0.5 (0.6) & 0.4 (0.5) \\
Freudenreich & 1.0 & 0.9 (1.4) & 2.0 (2.4) & 1.3 (1.6) & 0.7 (1.0) & 0.4 (0.5) \\
Dwek et al& 0.6 & 0.6 (0.9) & 1.2 (1.5) & 1.0 (1.1) & 0.5 (0.7) & 0.4 (0.6) \\
\hline
\end{tabular}
\end{center}
\caption{Optical depth towards $\ell = 3^{\circ}.9, b = -3^{\circ}.8$ 
in units of $10^{-6}$.  Recall that Popowski et al.'s (2000) value for
the optical depth to the red clump is $2.0 \pm 0.4 \times 10^{-6}$,
which should be compared to the numbers in the third column.  The
figures in parentheses include the effects of spiral structure. The
extra enhencement in the optical depth comes mainly from the extra
mass in the spiral models.}
\label{table:baade}
\end{table}
\clearpage
\begin{table}
\begin{center}
\begin{tabular}{l|cccccc}\hline
Direction & $\theta$ Mus & $\gamma$ Nor & $\gamma$ Sct & $\beta$ Sct\\
\hline
EROS & $< 0.68$ & $0.27^{+0.30}_{-0.17}$ & $1.64^{+0.92}_{-0.74}$ & $< 1.03$\\
\hline
Binney et al& 0.32 (0.56) & 0.48 (0.69) & 0.79 (1.07) & 0.60 (0.83)\\
Freudenreich & 0.47 (0.71) & 0.78 (1.18) & 1.11 (1.43) & 0.95 (1.23)\\
Dwek et al& 0.34 (0.61) & 0.51 (0.72) & 0.85 (1.13) & 0.64 (0.90)\\
\hline
\end{tabular}
\end{center}
\caption{Optical depth towards spiral arms in units of $10^{-6}$.
The figures in parentheses include the effects of spiral structure.
The numbers should be compared with the experimental data provided by
the EROS collaboration (Derue et al. 2001) in the second line.}
\label{table:spiral}
\end{table}
\clearpage

\end{document}